\documentclass[twocolumn,aps,prl]{revtex4}

\usepackage{amsmath}
\usepackage{amssymb}
\usepackage{times}
\usepackage{colordvi}
\usepackage{textcomp}

\begin{document}



\noindent
{\bf Z\"ulicke and Shimshoni Reply:}
In two recent articles~\cite{UZandESprl,UZandESprb}, we developed a
transport theory for an extended tunnel junction between two
interacting fractional-quantum-Hall edge channels, obtaining analytical
results for the conductance. Ponomarenko and Averin (PA) have
expressed disagreement with our theoretical approach and
question the validity of our results~\cite{PAcomment}. Here we show
why PA's critique is unwarranted.

The system, called a {\em line junction\/}, is formed by two 
counterpropagating single-branch quantum-Hall edges with different
fractional filling factors. It consists of three regions, as shown in Fig.~1 of
Ref.~\cite{UZandESprl}. A finite uniform tunneling amplitude exists only
within the central segment of length $L$, and a (screened Coulomb)
interaction {\em between\/} the two counterpropagating edge channels is
present {\em only in that region of space\/} as well. Outside of the central
region, tunneling and inter-edge interactions are switched off. However,
interactions between electrons {\em within\/} each edge channel are
considered to be finite everywhere. Our assumptions about the spatial
variation of tunneling and interactions are designed to provide a
realistic model of experimentally relevant situations~\cite{Expts}.

The inherent non-uniformity of the system under consideration makes
it necessary to implement proper matching conditions at the
interface between the central region and adjacent (lead) regions. This
is the crux of our disagreement with PA. Their results are based on the
assumption that the bosonic fields describing edge excitations in the
two counterpropagating branches should be continuous. This
approach has been successfully used to study tunneling transport
through point contacts between edge channels with {\em uniform\/}
interaction potential. However, PA's assumption will not hold for our line
junction because it ignores charging of the central region due to the
spatially inhomogeneous (piecewise-constant) inter-edge interaction.
Similar charging effects occur in a finite interacting quantum wire that is
attached to non-interacting one-dimensional leads~\cite{Egger,BlantButt}.
To properly account for these, a different set of boundary conditions needs
to be imposed.

Instead of the unphysical requirement of continuous bosonic fields (and,
hence, edge-charge densities), we demand continuity of a quantity we call
the {\em local chemical potential\/} $\mu_j$ for each edge mode. It is
defined as the functional derivative of the total system Hamiltonian w.r.t.\
the chiral edge density, $\mu_j(x) = \delta\mathcal{H}/\delta\varrho_j(x)$,
and is therefore the operator field that is canonically conjugate to
$\varrho_j$. A detailed discussion of this quantity's physical meaning can
be found, e.g., in Ref.~\onlinecite{SafiEurJ}. (See also related
articles~\cite{ElChemRel}.)
Its {\em expectation value\/} corresponds to an electrochemical potential
that would be established after electronic equilibration. The utility of the
operators $\mu_j$ arises from the fact that they are well-defined 
everywhere in the system, irrespective of whether and where equilibration
actually occurs. In particular, they are constant in the lead regions where
the chiral edge densities cannot change. Requiring continuity of the $\mu_j$
operators at an interface where electron interactions are discontinuous
amounts to implementing the proper self-consistent solution of the
equivalent of Poisson's equation for a short-ranged interaction
potential~\cite{Egger,BlantButt}.

Contrary to PA's claim, we {\em do not\/} impose equilibration in outgoing
edge-channel leads at their interface with the central line-junction
region. We demonstrate this fact by re-deriving our central result, relating
the junction current $I_{\text{J}}$ to the voltage drop [Eq.~(36) in
Ref.~\onlinecite{UZandESprb}], with explicit reference only to chemical
potentials $\mu_1$ and $\mu_3$ of {\em incoming\/} edge-channel leads.

Without loss of generality, we choose the chirality of edge channels in the
line junction as shown in Fig.~1b of Ref.~\onlinecite{UZandESprl}. The
local chemical potential of right-movers (left-movers) at $x=-L/2$ ($x=L/2$)
is fixed by the incoming edge-channel leads to be equal to $\mu_1$
($\mu_3$). Straightforward algebra yields
\begin{subequations}\label{boundChem}
\begin{eqnarray}
\mu_1 &=& \frac{2\pi\hbar}{\nu_{\text{R}}}\left\{ \sqrt{\tilde\nu} v_{\text{n}}
\varrho_{\text{n}}\!\left(\!-\frac{L}{2}\!\right) + \frac{I_{\text{J}}}{2} \right\} + 
\frac{2\pi\hbar v_{\text{c}} \varrho_{\text{c}} \! \left(\!-\frac{L}{2} \! \right)} 
{|\nu_{\text{R}}-\nu_{\text{L}}|} \, , \\
\mu_3 &=& \frac{2\pi\hbar}{\nu_{\text{L}}}\left\{ \sqrt{\tilde\nu} v_{\text{n}}
\varrho_{\text{n}}\!\left(\!\frac{L}{2}\!\right) - \frac{I_{\text{J}}}{2} \right\} + 
\frac{2\pi\hbar v_{\text{c}} \varrho_{\text{c}} \! \left(\! \frac{L}{2} \! \right)} 
{|\nu_{\text{R}}-\nu_{\text{L}}|} \, .
\end{eqnarray}
\end{subequations}
The charge mode has a trivial dynamics and satisfies $\varrho_{\text{c}}
(L/2)\equiv \varrho_{\text{c}}(-L/2)$. Hence we find the general relation
\begin{equation}\label{generalRes}
\frac{\mu_1-\mu_3}{2\pi\hbar} = \sqrt{\tilde\nu}\, v_{\text{n}}\left[ \frac{
\varrho_{\text{n}}\!\left(\!-\frac{L}{2}\!\right)}{\nu_{\text{R}}} - \frac{ \varrho_{\text{n}}\!\left(\!\frac{L}{2}\!\right)}{\nu_{\text{L}}}\right] +\frac
{\nu_{\text{R}}+\nu_{\text{L}}}{2\nu_{\text{R}}\nu_{\text{L}}} \, I_{\text{J}}\, .
\end{equation}
Some more algebra based on the continuity equation for the neutral mode
(see Sec.~III of Ref.~\onlinecite{UZandESprb}) yields the condition
$\varrho_{\text{n}}(L/2)= \varrho_{\text{n}}(-L/2)\equiv \bar\varrho_{\text{n}}$ 
for the stationary limit. With that, Eq.~(\ref{generalRes}) above specializes
to the quoted voltage-drop equation for the line junction, ostensibly derived
in  Ref.~\onlinecite{UZandESprb} by matching local chemical potentials at
interfaces. Here we find the same result without considering chemical
potentials in outgoing leads, thus confirming the accuracy of our approach.

Unlike in point contacts, an externally imposed voltage at a line junction
sustains both the current $I_{\text{J}}$ and a charge imbalance $\bar
\varrho_{\text{n}}$, which are related by the central region's
dynamics. PA's result for the conductance can arise only in the limit of
vanishing $\bar\varrho_{\text{n}}$, which we consider to be unphysical
because it implies the absence of an internal driving force for the current.

\vspace{1eX}

\noindent
U. Z\"ulicke \\
\indent Institute of Fundamental Sciences, Massey University, \\
\indent  Private Bag 11~222, Palmerston North, New Zealand 

\noindent
E. Shimshoni \\
\indent Department of Mathematics and Physics, University \\
\indent of Haifa at Oranim, Qiryat-Tivon 36006, Israel \\

\end{document}